\documentclass[twocolumn]{article}
\usepackage{xcolor}

\usepackage[noadjust]{cite}

\usepackage[T1]{fontenc}
\usepackage{textcomp}
\usepackage{optidef}
\usepackage{ulem}

\usepackage{stackengine}

\usepackage{amsthm}

\usepackage{enumitem}

\begin{document}
%
\title{Blockchain-Aided Flow Insertion and Verification in Software Defined Networks}


\author{Jiejun Hu, Martin Reed, Mays Al-Naday, Nikolaos Thomos\\
School of Computer Science and Electronic Engineering\\
University of Essex\\
Colchester, UK\\
Email: jiejun.hu@essex.ac.uk, mjreed@essex.ac.uk, mfhaln@essex.ac.uk, nthomos@essex.ac.uk
\thanks{This paper was presented in 2020 Global Internet of Things Summit (GIoTS) (pp. 1-6). IEEE.}}

\date{}

%



\maketitle

\begin{abstract}
The Internet of Things (IoT) connected by Software Defined Networking (SDN) promises to bring great benefits to cyber-physical systems.
However, the increased attack surface offered by the growing number of connected vulnerable devices and complex nature of SDN control plane applications could overturn the huge benefits of such a system. This paper addresses the vulnerability of some unspecified security flaw in the SDN control plane application (such as a zero-day software vulnerability) which can be exploited to insert malicious flow rules in the switch that do not match network policies.
Specifically, we propose a blockchain-as-a-service (BaaS) based framework that supports switch flow verification and insertion; and additionally provides straightforward deployment of blockchain technology within an existing SDN infrastructure. While use of an external BaaS brings straightforward deployment, it obscures knowledge of the blockchain agents who are responsible for flow conformance testing through a smart blockchain contract, leading to potential exploitation. Thus, we design a strategy to prevent the blockchain agents from acting arbitrarily, as this would result in what is termed a “moral hazard”. We achieve this by developing a novel mathematical model of the fair reward scheme based on game theory. To understand the performance of our system, we evaluate our model using a Matlab based simulation framework. The simulation results demonstrate that the proposed algorithm balances the needs of the blockchain agents to maximise the overall social welfare, i.e. the sum of profits across all parties.
\end{abstract}



\section{Introduction}
Software Defined Networking (SDN) has been proposed to increase network scalability and improve flexibility, through decoupling of the data and control planes.
Such characteristics have rendered SDN as an attractive choice for connecting Internet of Things (IoT) use-cases, particularly those cannot be supported by legacy networks \cite{Bera17,Ramos19}. However, the inherit trust of the switch to the controller and the application running in the controller leaves the SDN infrastructure vulnerable to a cyber-attack through which malicious flow rules may be inserted into the SDN switch(es)\cite{Eom19}.
Here, we consider two security problems in SDN: first, there is no \textit{flow conformance test} applied when updating the flow table of a switch, i.e. checking that a flow conforms to some external policy; second, attackers may attempt to tamper with  messages between the controller and the switch. Either of these scenarios would allow an attacker to seriously undermine the data-plane; for example, by diverting traffic for malicious inspection or blocking services to perform denial of service and these are serious concerns for operators that are deploying SDN.

To address the security issues above, this paper proposes a novel solution which carries out an independent confirmation of the SDN control plane action. The use of an independent confirmation is important as it can be separate from the underlying SDN application, which may be vulnerable. This is performed through a flow conformance test implemented within a smart contract of a blockchain (BC) \cite{hu2020blockchain}. BC technology, a sub-concept of distributed ledger technology, is essentially an append-only data structure maintained by a group of \textit{not-fully-trusted} nodes, that nevertheless provide a \textit{trusted} data structure through a suitable consensus algorithm \cite{dinh2017blockbench}. Furthermore, BC is decentralized, immutable, transparent, and reliable, which allows it to stand independently from the SDN network and the IoT network to establish a distributed trust mechanism. The mechanism described within this paper should be seen as one part of a wider security architecture. An example of such a security architecture for IoT systems is that proposed by the SerIoT project~\cite{gelenbe2018security} which combines novel network security mechanisms~\cite{amangele2019hierarchical,mitev2019man} together with wider analytic strategies and tests them in real-world use cases.

The work is inspired by recent studies that have proposed different architectures to establish blockchain-based SDN scenarios. 
A novel blockchain-based distributed cloud architecture has been proposed in \cite{sharma2017software,sharma2018softedgenet}, with fog nodes at the edge providing functionality of the SDN controller.
Their work introduces a hierarchy with a central blockchain-based cloud moving towards a blockchain-based edge, with the latter taking responsibility of updating the flow rules.
Although the works in \cite{sharma2017software,sharma2018softedgenet} show how cloud/fog computing would support blockchain, the solution neglects the issue of flow conformance testing that we address in this paper.
Studies\cite{muthanna2019secure,rathore2019blockseciotnet} also attempt to solve the security challenges in IoT by using blockchain to dynamically update an attack detection model. Boukria et al.\cite{boukria2019bcfr} propose a blockchain-based controller against false flow rule injection. The authors mainly focused on the SDN controller authentication.  
However, these solutions fail to exploit blockchain technology in a manner that is compatible with existing SDN architectures. Specifically, they do not consider how to verify a new flow using blockchain in fine-grained manner; nor, how to enable blockchain technology without changing the foundation of SDN. 
This paper aims to solve these problem by introducing the novel security solution for blockchain-based SDN (BC-SDN). We do this through designing a flow conformance testing workflow that uses smart contracts and solve the reward scheme through a novel mathematical analysis.

\subsection{Architecture of Blockchain-based SDN}
\begin{figure}[tb]
\centering
\includegraphics[width=2.5in]{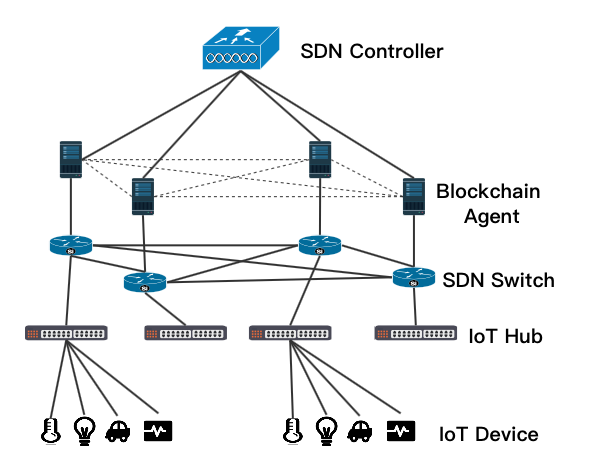}
\caption{System model of blockchain-based SDN}
\label{fig: archv1}
\end{figure}
We consider an architecture that consists of traditional SDN and Blockchain-as-a-Service (BaaS) model \cite{samaniego2016blockchain}, in a generic IoT scenario. In this scenario, IoT devices communicate with each other through a SDN-based network\cite{gelenbe2018security}. The communication is realised through data-plane paths (and actions) that result from a centralised SDN application in the central SDN controller; this application inserts flow rules in the SDN switches according to some predetermined algorithm. It is this SDN application, or the controller itself, which is open to vulnerabilities that may allow an attacker to cause malicious behaviour. We incorporate a blockchain overlay to act as an intermediary verification plane between the control and forwarding counterparts to stop malicious flow insertion. Fig. \ref{fig: archv1} shows an architectural view of the entities in our BC-SDN proposition, including:
a) \textbf{IoT devices} that can sense the environment, upload sensory data and control actuators; b) \textbf{IoT hubs} that connect the IoT devices to the SDN-switches; c) \textbf{SDN switches} that detect new flows and execute a forwarding plan calculated by the \textbf{SDN controller(s)}; d) \textbf{Blockchain agents (BCAs)} that are servers provided by a third party. BCAs are in charge of flow conformance testing, which includes flow verification and validation via smart contract. Furthermore, BCAs also execute basic blockchain functions, such as consensus process, sending transactions, and maintaining the shared ledger. And finally e) the \textbf{Controller} with a global view of the network and which calculates the best path of the packets according to some pre-defined policy.

In this paper we adopt a permissioned-based consortium blockchain, such as Hyperledger, which means that only authorized entities can conduct the blockchain functions. Futhermore, we adopt BaaS infrastructure with a BCA colocated with an SDN switch. There are three main advantages of using the proposed mechanism for flow conformance testing.  First, as BCAs are colocated with the SDN switches, providing a blockchain service to the SDN, the BCAs confirm the secure communication between the controller and the corresponding switch. Second, BCAs are components provided by an independent entity, which conducts flow verification/validation outside the SDN network to guarantee the connection privacy. Finally, BaaS enables straightforward deployment of blockchain in SDN without the SDN operator needing to create their own blockchain system. 

\subsection{Solution and Contribution}
In the proposed architecture, 
BCAs verify the conformance of a flow as an external service, while the SDN network entities are the \textit{users} of the BCA service. Thus, a reward strategy is critical to stimulate the BCAs to perform the verification and validation action. When a leading BCA initializes the flow conformance testing procedure, it hires a group of BCAs as verifiers and offers them a reward. The verifiers have knowledge of: their computation ability, channel condition, and the size of the task. Hence, the verifiers can decide how much reward they want. However, the verifier may behave greedily and demand a higher price for a simple task, or else they will not implement the task. This phenomenon is known as a "moral hazard" of the verifiers. The underlying reason for the moral hazard is the asymmetric relationship about system information, i.e. the verifiers have more information to make a decision than the leading BCA. To maintain the execution of flow verification/validation, the leading BCA needs a new reward scheme to overcome the moral hazard of verifiers.

The design of the reward mechanism is critical in a cyber-physical system and game theoretic methods are widely used in solving this problem. Due to the fact that blockchain technology is a resource consuming application, studies concentrate on optimisation problems that consider various aspects such as: the latency, throughput of transactions, finality, security, and decentralization\cite{kang2019toward,liu2019performance}. In this paper, to solve the moral hazard of verifiers, we analyze and quantify the verification/validation latency which is related to the blocksize of the BCAs. Then, we design the reward mechanism based on contract theory\cite{bolton2005contract}. The main contributions of this paper are summarized as follows:
\begin{itemize}[leftmargin=*]
\item We first propose a Blockchain-aided Software-Defined Network, a novel framework that uses blockchain technology to provide flow conformance testing in SDN. 
\item We design the workflow of flow insertion and verification of BC-SDN with the assistance of blockchain.
\item We devise a fair reward scheme based on contract theory to stimulate the BCAs.
\end{itemize}
To the best of our knowledge, our work is the first to propose the workflow of flow rule verification in BC-SDN, and also to use contract theory to study the quantified performance of blockchain in this use-case. 

\section{Workflow of BC-SDN}
\begin{figure}[tb]
\centering
\includegraphics[width=3.0in]{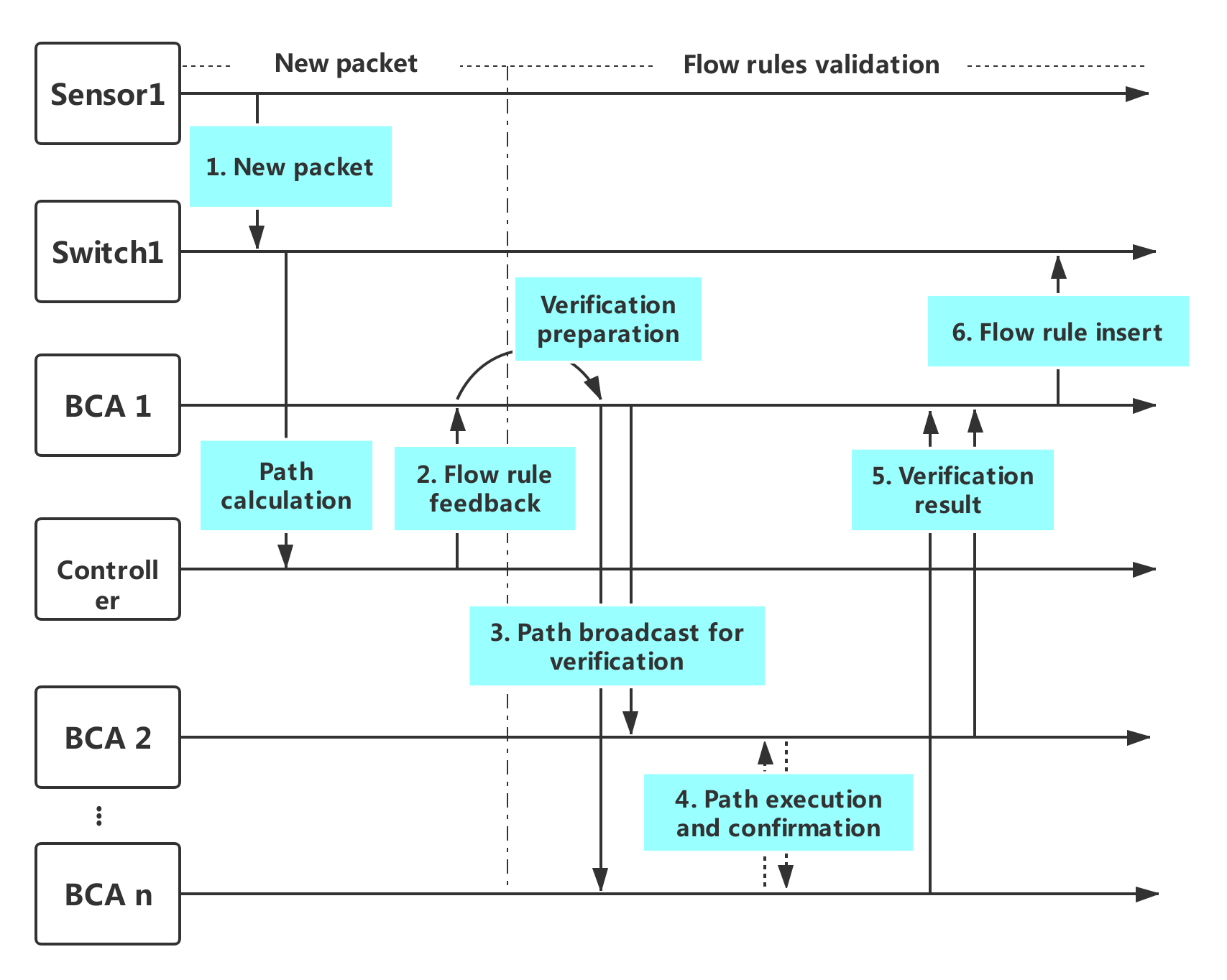}
\caption{Workflow of BC-SDN}
\label{fig: flow}
\end{figure}

In order to explain the proposed workflow of BC-SDN we use Fig. \ref{fig: flow}, which presents a sample network. We consider that Sensor 1 generates a new packet. First, we assume that:
\begin{enumerate}
\item This blockchain application is using one unique Hyperledger \textit{channel}, although other users may be using different \textit{channels}.
\item The IoT devices/IoT hubs that require new flow rules have been registered and enrolled with the organization's Certificate Authority and received back necessary cryptographic material, which is used to authenticate a device.
\item The BCAs have been fed with previous topology and connectivity information from the controller.
\item The controller is responsible for path calculations, which will result in a set of flow rules according to the path. 
\item All communication channels are secured through a mechanism such as transport layer security (TLS) authenticated using appropriately deployed certificates.
\item The blockchain implements a next generation smart contract which keeps all data secure, as described by Cai et al.~\cite{Cai19}
\item The switch and the controller use an unmodified OpenFlow protocol so that the validation is performed as a transparent process; any validation failure is communicated using standard OpenFlow error messages.
\end{enumerate}

The actions of establishing and testing conformance of a new flow rule are embedded in a smart contract within the blockchain. The endorsement policy states that it requires $n$ verifiers to establish a new flow rule. Below, we explain these actions in more depth.
\begin{enumerate}
\item Sensor1 causes Switch1 to initiate a new flow establishment request when it sends a packet with source IP/port information, destination IP/port information, and protocol. The message is formed as \textit{packet = <souIP, desIP, souPort, desPort, Protocol>} without an existing flow rule, and is sent to the controller over the secure communication channel. The controller calculates the path according to the \textit{packet} and generates new flow rules that it forwards to the corresponding BCA (verification initiator, VI). The ID of flow is \textit{fid= hash(packet)}. This request of the controller targets BCA1. The VI constructs a new flow proposal \textit{<PROPOSAL, tx, Consig>} ensures the new flow proposal is properly formed; \textit{tx} includes: the ID of controller, ID of the flow, packet, ID of smart contract, endorsement policy, and time stamp. The VI's credentials are used to produce a signature \textit{ConSig = hash(tx)} for this proposal. 
\item The VI starts the verification of the new flow by employing verifier peers. The verifiers check that: the proposal is well formed; it has not been submitted already in the past; the signature is valid; and that the VI is properly authorized to perform the proposed operation. The verifiers use the new flow proposal as input to invoke the smart contract. The smart contract checks the new flow against the flow conformance policy and asserts as TRUE or FALSE accordingly. All the response values will be stored in blockchain's status database \textit{readset} and \textit{writeset}. The response values along with the verifiers' signature is passed back to the VI as a "proposal response" \textit{ProRes=<TransactionEndorsed, fid, TranProposal, epSig>}, where \textit{TranProposal = (epID, fid, chaincodeID, tx, readset, writeset)}. If this new flow is invalid, then send message \textit{<TransactionEndorsed, fid, REJECT, epSig>}. Note that no changes are made to the flow ledger till now.
\item Proposal responses are checked as follows. The VI inspects the verifiers' signature and confirms that the number of identical \textit{ProRes} responses reach the number expected by the endorsement policy. If the VI only enquires of data from the ledger, then there is no need to update the blockchain database, the \textit{ordering service}. The VI checks the endorsement policy has been reached before storing the new flow in the ledger.
\item Ordering service: a service in Hyperledger implemented in a separate group of peers called \textit{orderers} that preform transaction ordering and maintain the distributed ledger. The VI broadcasts the proposal \textit{<PROPOSAL>} with \textit{<ProRes>} in one message \textit{broadcast =(PROPOSAL, ProRes)} to all of the orderers in the ordering service. The leader of the ordering service calls the ordering peers by sending the message \textit{deliver(seqno, prevhash, endorsement)}, where \textit{seqno} is sequence number, \textit{prevhash} is the hash of the most recently delivered endorsement. The ordering service orders the messages chronologically and creates blocks of transactions.
\item New flow validation and committing: The blocks of transactions are delivered to all BCAs. The BCAs verify the ID of smart contract, endorsement policy, and consistency of the status database to avoid violations. If the checks pass, the transaction is deemed valid or committed. In this case, the BCAs set the bitmask of the \textit{PeerLedger}. If the checks fail, the new flow establishment is considered invalid and the BCA unsets the bitmask of the \textit{PeerLedger}. This invalid transaction is stored until it is deleted by a periodic blockchain function.
\item Ledger updated: Each BCA appends the block with the new \textit{fid}. At the same time, the VI is notified that the new flow has been immutably appended to the chain.
\end{enumerate}

\section{System model}
We use BaaS in BC-SDN, i.e. we rent the servers from a third party.  This means that the VI lacks of knowledge of the performance (and other abilities) of the servers, thus there is asymmetry between the knowledge of the VI and the BaaS. This information asymmetry leads to what is known as a moral hazard\cite{holmstrom1979moral}. The moral hazard is commonly solved by contract theory in field of economics. Motivated by the above, in this paper we design a reward scheme of BaaS, which can solve the moral hazard of the third-party BCAs. In BC-SDN, flow conformance testing is provided by BCA verifiers and a contract is designed based on the outcome of the verifiers. The verifier offers two flow conformance testing plan. A simple flow conformance testing that only verifies the source and destination MAC address and Port number. And a complex flow conformance testing that verifies the whole path. When a flow conformance initiates, the VI presents a contract for the verifiers offering a reward. Then, the verifiers have the option to either accept the reward or refuse. According to the reward, the verifiers exert the flow conformance testing plan. In this section, we first introduce the system model by analysing the cost and income of the VI and the verifiers. Then, we define the employed utility functions of both the VI and the verifiers. Finally, we propose our solution based on contract theory.

\subsection{Execution cost of a verifier} 

Verifier $i\in \mathcal{V}=\{1, ..., n\}$ who participates in verification can select the blocksize from a set $\mathcal{S} = \{s_1, ..., s_n\}, n \ge 1$. We define verifier $i$ select blocksize $s_i \in \mathcal{S}$ Similarly to \cite{hu2020blockchain}, the execution cost of a verifier is defined in quadratic form, which is thus convex and provides a straightforward evaluation of the derivative. 
\begin{equation}
\centering
\phi(s_i) = \frac{1}{2} \alpha s_i^2, \quad s_i \in \mathcal{S} 
\end{equation}
where $\alpha$ is a parameter that reflects the cost factor of a verifier.

\subsection{Performance evaluation of a verifier}
In contract theory, the VI can only observe the performance of the verifier but not the real effort it exerts. Consider a blocksize $s_i$ the verifier exerts is invisible from the VI, but the performance of the verifiers is observable. In this paper, the performance metric is the verification/validation latency, and is defined as $\mathcal{L} = \{l_1, ..., l_n\}, n\ge1$. We assume the latency is proportional to the blocksize defined as $l_i = f(s_i)$, where function $f(\cdot)$ is monotonically increasing with $s_i$. For the sake of simplicity, we assume a linear relation as $l_i = s_i $, without affecting the generality of our conclusions. Note that we assume that the latency in the Hyperledger fabric iteself is the same for all users and is thus not considered here.

\subsection{Reward plan for a verifier}
The reward for the verifier $i$ is denoted as $\mathcal{R} = \{r_1, ..., r_n\}, n\ge1$. $r_i \in \mathcal{R}$ means the reward for per unit of performance. We define the income for the verifier as $r_i l_i$. In this model, the VI considers only the latency that the verifier provides and wishes to incentivise the verifier to provide the more complex flow conformance test.  

\subsection{Utility functions of VI and verifier}
In our model, the VI considers only the latency of the verifier because it expects the verifiers perform complex flow conformance testing, which leads to long latency. We define the utility function of VI as the gross benefit minus the reward to the verifier. Thus, the VI's utility function is written as
\begin{equation}
\centering
U_v =   \beta l_i - r_i l_i 
\label{eq: u_v1}
\end{equation}
where $\beta$ is the income factor of VI and $r_i l_i$ is the reward plan for a verifier. VI needs to enquire the latency of verifier to offer the reward. So, we assume that the VI has full trust in the latency verifier reports. Thus, we substitute $l_i$ with $s_i$, the utility of VI can be rewritten as
\begin{equation}
\centering
U_v = \beta s_i - r_i s_i
\label{eq: u_v2}
\end{equation}
The utility of the verifier considers the reward plan minus the execution cost. Thus, the utility of verifier $i$ is defined as
\begin{equation}
\centering
U_i =  r_i s_i -\frac{1}{2} \alpha s_i^2
\label{eq: u_vi}
\end{equation}
\subsection{Social Welfare}
We define the social welfare $\omega$ of the BaaS service as the profit of the verifier and the VI. Thus, from the utility of the verifier (\ref{eq: u_v2}) and the utility of the VI (\ref{eq: u_vi}), the social welfare is
\begin{equation}
\centering
\omega =  U_i + U_v = \beta s_i - \frac{1}{2} \alpha s_i^2 
\label{eq: sw}
\end{equation}

\section{Problem formulation}
As mentioned earlier, a contract is designed as the tuple $(\mathcal{R}, \mathcal{L})$, as the VI rewards the verifiers according to the latency. The VI aims to use the minimal reward to obtain the optimal blocksize $s_i^*$ to guarantee the flow verification/validation is in progress. Meanwhile, the verifier chooses the optimal blocksize according to the reward. According to our system model, we can formulate the maximization of the verifier's utility as
\begin{align}
&\max_{s_i}  r_i s_i -\frac{1}{2} \alpha s_i^2\label{eq: p1.1}\\
&\text{s.t.} \quad r_i s_i  -\frac{1}{2} \alpha s_i^2 \ge \sigma\\
& \quad \quad s_i < s_{max}
\end{align}
where $s_{max}$ is the biggest blocksize. And the maximization problem of VI is formulated as
\begin{align}
&\max_{r_i, s_i^*} \beta s_i - r_i s_i \label{eq: p1.2}\\
&\text{s.t.} \quad \beta s_i - r_i s_i\ge \sigma \label{eq: p1.2_1}
\end{align}
where $\sigma$ is the reservation utility of verifier. Reservation utility is the minimum profit that must be guaranteed by the contract to make it acceptable to the verifiers. The optimization problem aims to maximize the utility of VI by proposing a reward plan according to the latency of a verifier. Through solving the maximization problem (\ref{eq: p1.1}) we determine the optimal blocksize $s_i^*$, which is then used to maximize the utility of verifier. The maximization problem (\ref{eq: p1.2}) guarantees that under the optimal blocksize, the VI maximizes its utility by choosing the optimal reward $r_i^*$.

For simplicity, but without loss of generality, we assume only one VI and a group of verifiers $\mathcal{V}$. The verifier $i$ offers two different verification plans. We denote $d$ as the verification plan indicator, where $d \in \{\mu,\iota\}$. When $d = \mu$, the verifier $i$ will preform a complex verification with a larger blocksize $s_i^\mu$, which has a longer latency $l^\mu$. On the contrary,  $d = \iota$ indicates a simple verification that verifies the source and destination IP address with a smaller $s_i^\iota$, which has the smaller latency $l^\iota$. An example of a complex verification could be a whole path of a flow, as opposed to a simpler verification which could be only the source and destination addresses.

For a verification task, the verifier with probability $p=p^\mu$ chooses the larger  blocksize $s_i^\mu$ and with probability $p^\iota$ chooses a smaller blocksize $s_i^\iota$, where $p^\iota= 1-p$ . We assume that $s_i=s_i^\mu = \varepsilon s_i^\iota$, where $\varepsilon \in (0,1)$ is blocksize difference factor. We can reformulate the optimization problem (\ref{eq: p1.1}) and (\ref{eq: p1.2}) in two stages: the maximization problem of verifiers and the maximization problem of VI by substituting the optimal verifier's blocksize $s_i^*$. The first stage of the optimisation problems is defined as
\begin{align}
&\max_{s_i} p(r_i s_i -\frac{1}{2} \alpha s_i^2) + (1-p)(\varepsilon r_i s_i -\frac{1}{2} \alpha \varepsilon^2 s^2) \label{eq: p1.1_simple}\\
&\text{s.t.} \quad s_i< s_{max} \label{eq: ps_1.1}
\end{align}
First, we find the derivative of the objective function in (\ref{eq: p1.1_simple}) and set it equal to zero to find the critical points of the function.  Hence, we obtain 
\begin{align}
\centering
 \frac{\partial U}{\partial s_i} &=p(r_i  - \alpha s_i) + (1-p)(\varepsilon r_i  - \alpha \varepsilon^2 s_i)=0 \nonumber \\
 s_i^* &= A r_i
\end{align}
where $A = \frac{p+(1-p)\varepsilon}{p\alpha+(1-p)\alpha \varepsilon^2}$. The second stage of the maximization problem is defined as following
\begin{maxi!}|s|[3]                   
    {r_i, s_i^*}                               
    { p(\beta s_i^*- r_i s_i^*) + (1-p)(\varepsilon\beta s_i^* - \varepsilon r_i s_i^*) \label{eq: p1.2_simple}}   
    {\label{eq: opt_s}}             
    {}                                
    \addConstraint{p(r_i s_i^* -\frac{1}{2} \alpha {s_i^*}^2) + (1-p)(\varepsilon r_i s_i^* -\frac{1}{2} \alpha \varepsilon^2 {s_i^*}^2)}{\ge \sigma \label{eq: ps_2.1}} 
\end{maxi!}
By substituting the optimal $s_i^*$ in (\ref{eq: p1.2_simple}) and (\ref{eq: ps_2.1}), we redefine the problem as
\begin{align}
&\max_{r_i, s_i^*}p(\beta A r_i - A r_i^2) + (1-p)(\varepsilon\beta A r_i - \varepsilon A r_i^2)  \label{eq: p1}\\
&\text{s.t.} \quad p(A r_i^2 -\frac{1}{2} \alpha A^2 r_i^2) + (1-p)(\varepsilon A r_i^2 -\frac{1}{2} \alpha \varepsilon^2 A^2 r_i^2)= \sigma \label{eq: ps_2'}
\end{align}
Similarly, by solving the first stage of the optimization problem, we determine the optimal blocksize, $s_i^*$, and the optimal reward, $r_i^*$, which are given by 
\begin{align}
\centering
r_i^* &=\frac{p\beta+(1-p)\varepsilon\beta}{p\alpha A+ (1-p)\alpha\varepsilon^2A} \label{eq: optimal_r}\\
s_i^* &=\frac{p\beta+(1-p)\varepsilon\beta}{p\alpha+ (1-p)\alpha\varepsilon^2} \label{eq: optimal_s}
\end{align}

\section{Simulation}
In this section, we numerically evaluate the proposed reward scheme in BC-SDN. We assume that the BCAs work as verifiers/validators. The latency includes the verification and validation latency. We outsource the committing and ledger updating to  GoLevelDB\cite{thakkar2018performance}, thus the latency of committing and ledger updating can be neglected. 

In the simulation set up, we assume that the reservation utility of VI and verifier $\sigma =0$. The reason we set the reservation utility is that, the optimal blocksize and reward remain the same, no matter how $\sigma$ changes (\ref{eq: optimal_r}) and (\ref{eq: optimal_s}). We demonstrate the impact of the income factor $\beta$, the cost factor $\alpha$, the probability $p$ of selecting a bigger blocksize, and the difference, $\varepsilon$, between the two blocksizes. For the sake of comparison, we compare the proposed reward scheme with a solution based on the Stackelberg game \cite{hu2018reward} which assumes the VI knows all the information of verifier where optimal blocksize will exert according to the optimal reward.

\begin{figure}[tb]
\centering
\includegraphics[trim={0 0 0 8mm},clip,width=3.0in]{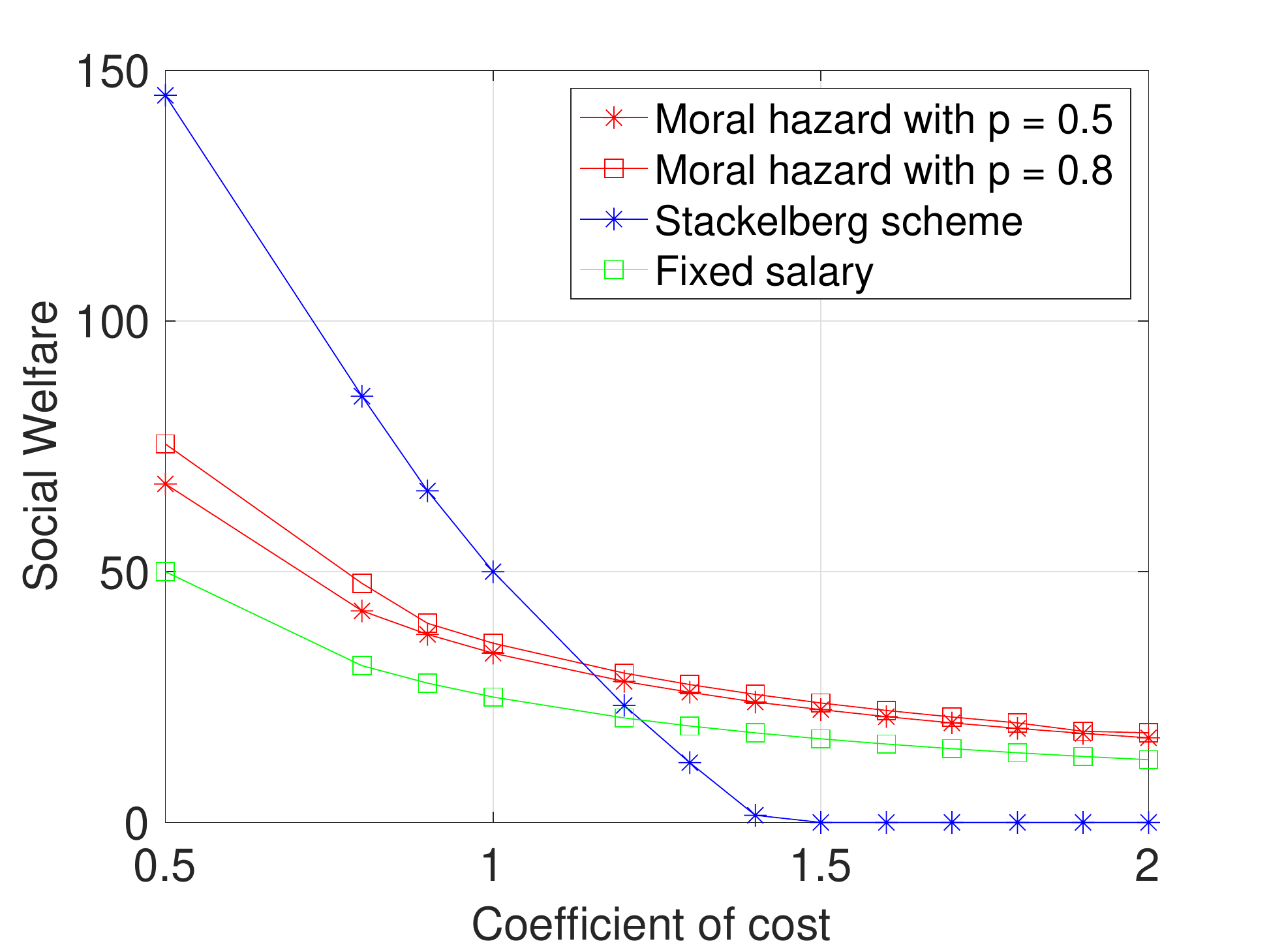}
\caption{Social welfare with respect to Cost factor $\alpha$}
\label{fig: simu_1}
\end{figure}

In Fig. \ref{fig: simu_1}, we compare the social welfare of three reward mechanisms as it varies with the cost factor $\alpha$ and the probability of selecting a bigger blocksize $p$. From the simulation results, we can see as $\alpha$ increases, the social welfare decreases. The reason for this behaviour is that the larger cost factor, the larger the cost of the verifier, which means that it is going to cost more to maintain the same blocksize. However, the proposed reward mechanism achieves higher social welfare when $\alpha$ is about 1.3. This is due to the fact that in the moral hazard scenario, the verifier can choose a smaller blocksize to compensate the execution cost in order to maintain the social welfare.

\begin{figure}[tb]
\centering
\includegraphics[trim={0 0 0 8mm},clip,width=3.0in]{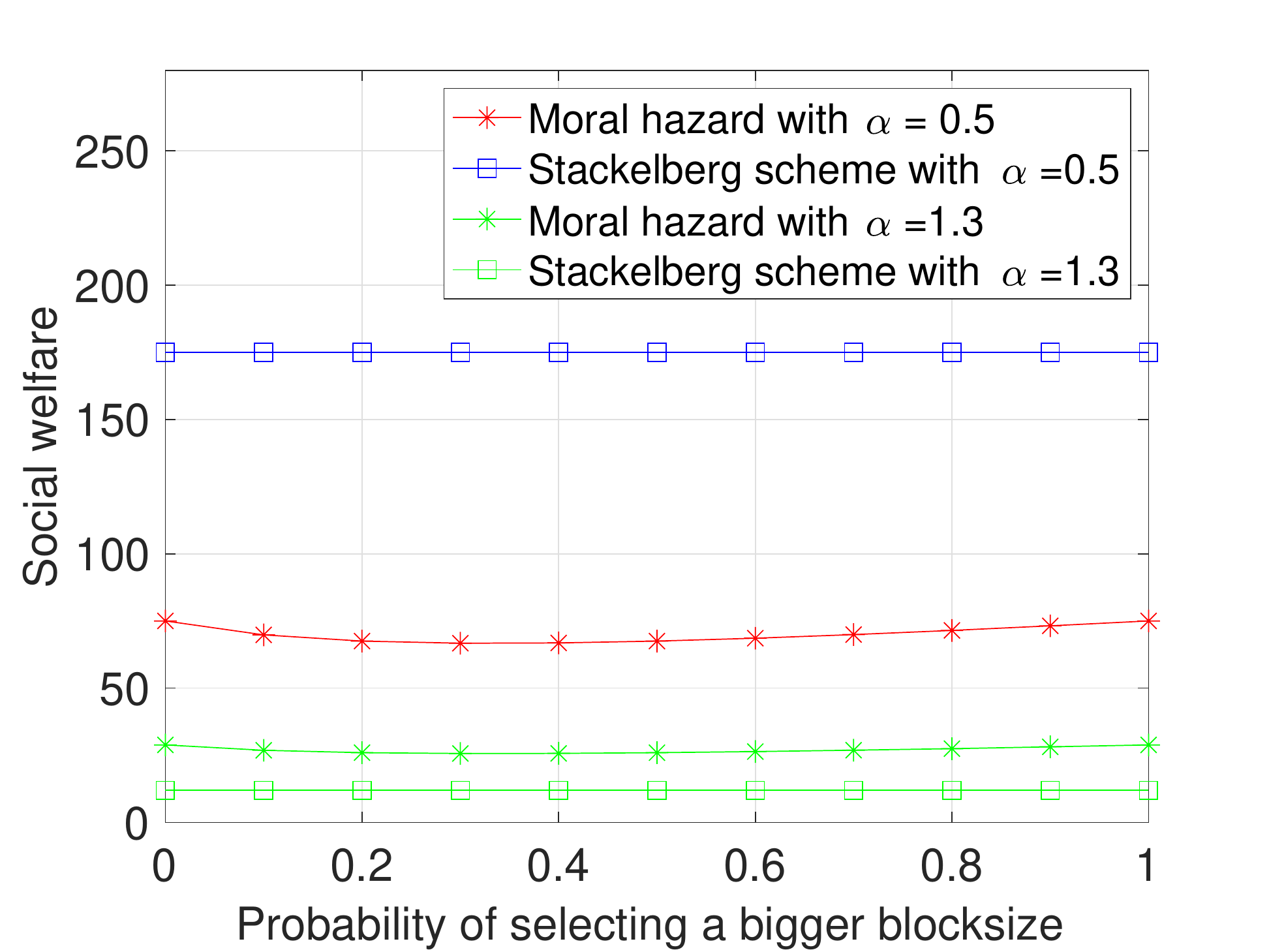}
\caption{Social welfare with respect to Probability of selecting a bigger blocksize for different $\alpha$ values}
\label{fig: simu_2}
\end{figure}

In Fig. \ref{fig: simu_2}, we analyse the impact of the probability of selecting the bigger blocksizes on social welfare. From the evaluation, it is obvious that there is only one optimal blocksize in the Stackelberg scenario. From this figure, we also note that for different cost factors, the proposed scheme can maintain the social welfare with a higher cost.

\begin{figure}[tb]
\centering
\includegraphics[trim={0 0 0 8mm},clip,width=3.0in]{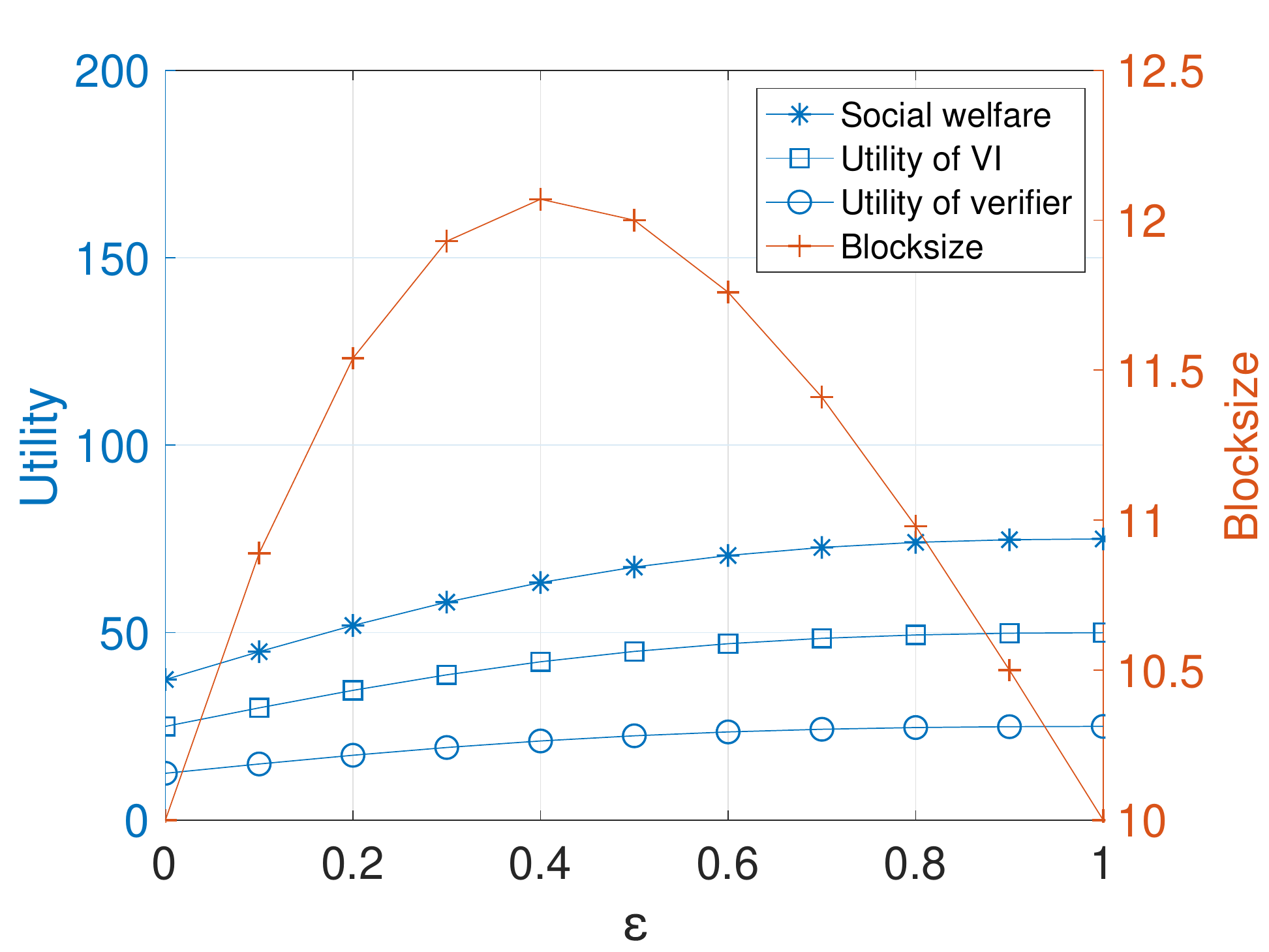}
\caption{Probability and social welfare}
\label{fig: simu_3}
\end{figure}

In Fig. \ref{fig: simu_3}, we evaluate the impact of the difference in the blocksizes with different values of $\varepsilon$. Note that when $\varepsilon$ tends to $0$, i.e., the difference between the blocksizes $s^\mu$ and $s^\iota$ is bigger. When it is $\varepsilon=1$, the blocksize is the same. For a neutral setting with $p=0.5$, $\alpha =0.5$, and $\beta = 10$, we observe that the utilities of the verifier and the VI increase when the blocksize's difference is increasing but that the rate of this increase reduces as the blocksizes tend towards the same size. At the same time, the biggest blocksize is observed as when $\varepsilon = 0.4$. The reason is that the optimal blocksize is only related with $p$ and $\varepsilon$ in (\ref{eq: opt_s}). 

\begin{figure}[tb]
\centering
\includegraphics[trim={0 0 0 8mm},clip,width=3.0in]{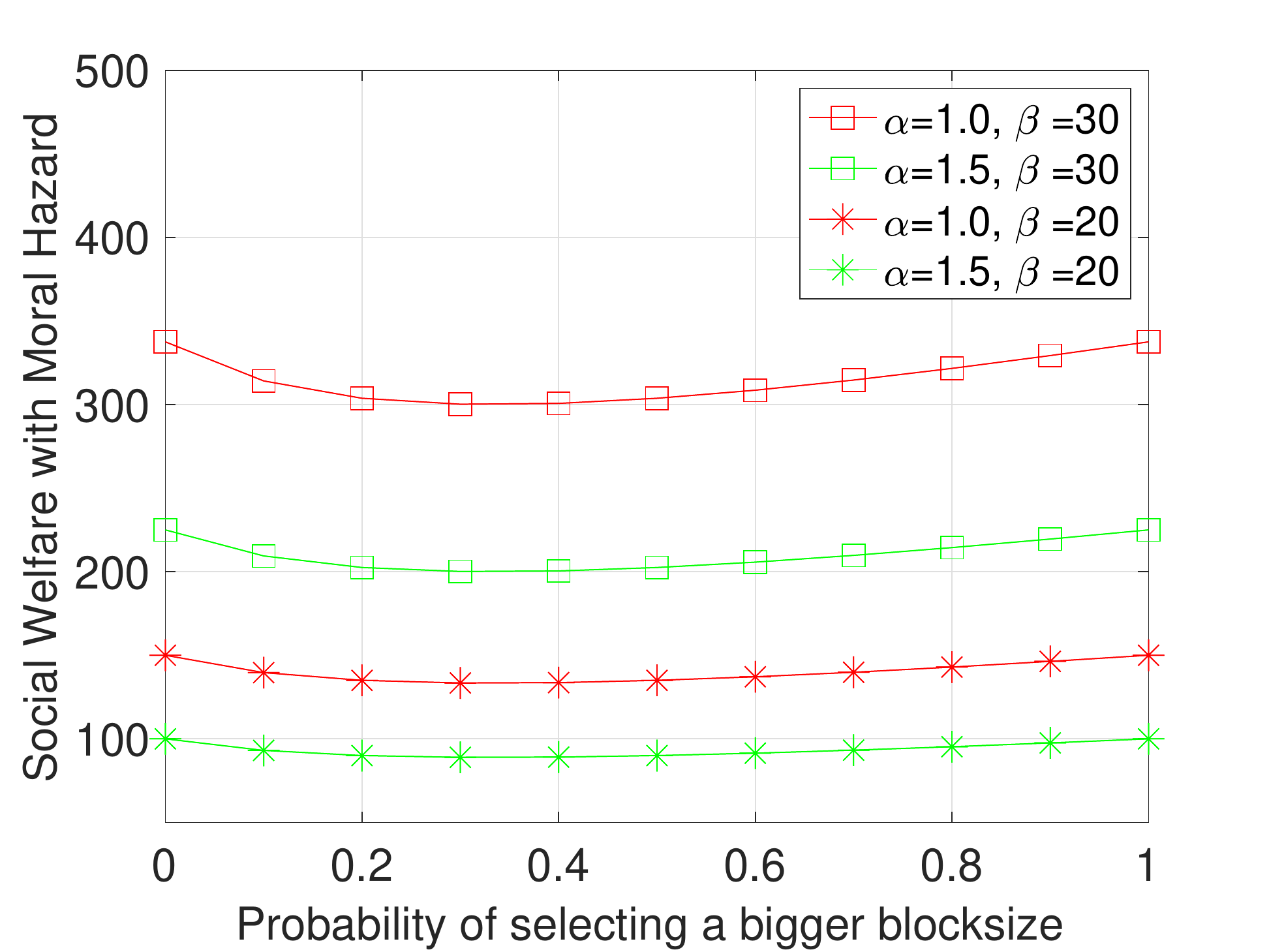}
\caption{Social welfare with respect to the probability of selecting a bigger blocksize for various $\alpha$ and $\beta$ values}
\label{fig: simu_4}
\end{figure}

To understand the impact of the income factor $\beta$ of VI and the cost factor $\alpha$ on the social welfare in the moral hazard scenario, we conduct a simulation  with results shown in Fig. \ref{fig: simu_4}. We can observe that if the income factor is bigger, the social welfare always increases irrespective of the probability $p$. Under the same income factor, the the bigger cost factor is, the smaller the social welfare. Also, we can observe that the lowest point of the social welfare appears when $p=0.4$. 

\section{Conclusion}
In this paper, we have investigated a novel security solution of SDN by adopting blockchain technology. According to the architecture of BC-SDN, we have proposed the workflow of new flow verification and insertion. Owing to the fact that we use BaaS, we have designed a reward scheme to tackle the potential moral hazard caused by the BCAs from the third party. By using the proposed reward scheme based on contract theory, we obtain the optimal blocksize of the blockchain and the corresponding reward for it. Finally, we evaluate our system to demonstrate the impact of different parameters using two different incentive mechanisms. The results show that the proposed reward scheme can gain good social welfare when the cost factor is high compared to Stackelberg reward scheme.

\section*{Acknowledgment}
This work was carried out within the project SerIoT, funded by the European Union's Horizon 2020 Research and Innovation programme under grant agreement No 780139.

\normalem
\bibliography{conf_v2.bib}
\bibliographystyle{IEEEtran}


%

\end{document}